\documentclass[11pt]{article}
\usepackage{mathrsfs}
\usepackage[centertags]{amsmath}
\usepackage{amsfonts}
\usepackage{amssymb}
\usepackage{amsthm}
\usepackage{cases}
\usepackage{indentfirst}
\usepackage{epsfig}
\usepackage {graphicx}
\usepackage{color}
\usepackage{CJK}

\date{}

\newtheorem{theorem}{Theorem}[section]
\newtheorem{corollary}{Corollary}[section]
\newtheorem{lemma}{Lemma}[section]

\newtheorem{remark}{Remark}[section]
\newtheorem{definition}{Definition}[section]
\numberwithin{equation}{section}

\allowdisplaybreaks[4]

\begin{document}
\title{\textbf{New sufficient conditions of signal recovery with tight frames via $l_1$-analysis}\thanks{
{Corresponding author. E-mail address: wjj@swu.edu.cn (J. Wang).} }}
\author{Jianwen Huang$^a$,\quad Jianjun Wang$^a$,\quad Feng Zhang$^a$,\quad Wendong Wang$^b$\\
{\small $^a$School of Mathematics and Statistics, Southwest
University, Chongqing, 400715, China}\\{\small $^b$School of Computer and Information Science, Southwest
University, Chongqing, 400715, China}\\{\small}}

\maketitle
\begin{quote}
{\bf Abstract.}~~The paper discusses the recovery of signals in the case that signals are nearly sparse with respect to a tight frame $D$ by means of the $l_1$-analysis approach. We establish several new sufficient conditions regarding the $D$-restricted isometry property to ensure stable reconstruction of signals that are approximately sparse with respect to $D$. It is shown that if the measurement matrix $\Phi$ fulfils the condition $\delta_{ts}<t/(4-t)$ for $0<t<4/3$, then signals which are approximately sparse with respect to $D$ can be stably recovered by the $l_1$-analysis method. In the case of $D=I$, the bound is sharp, see Cai and Zhang's work \cite{Cai and Zhang 2014}. When $t=1$, the present bound improves the condition $\delta_s<0.307$ from Lin et al.'s reuslt to $\delta_s<1/3$.

\end{quote}

{\bf AMS Classification(2010)}:~~94A12, 94A15, 94A08, 68P30, 41A64, 15A52, 42C15

\section{Introduction}
\label{sec1}
In recent years, compressed sensing (CS) has brought about important research activity. Formally, in CS, researchers are interested in the model below
\begin{align}\notag
b=\Phi f+z,
\end{align}
where $\Phi\in\mathbb{R}^{M\times N}$ $(M\ll N)$ is a known measurement matrix, observed signal $b\in\mathbb{R}^M$, an unknown signal $f\in\mathbb{R}^N$ and $z\in\mathbb{R}^M$ is a vector of measurement errors. In the standard compressed sensing frame, if the signal is (nearly) sparse in the nature base or other orthonormal bases, it can be accurately or stably recovered in the noiseless case or the noise case respectively under various sufficient conditions on the sensing matrix $\Phi$, for instance a restricted isometry property (RIP) condition, see [1-9].

However, in practical applications, there exist a large number of signals of interest whose sparsity is not expressed in terms of an orthogonal basis. On the contrary, these signals are sparse with respect to an overcomplete dictionary or a tight frame \cite{Bruckstein et al,Candes et al} or a general frame \cite{Liu et al}. That is to say, the original signal $f\in\mathbb{R}^N$ can be represented as $f=Dx$ where $D$ is some coherent and redundant dictionary with the size $N\times d$ $(N\leq d)$ and $x$ is (approximately) sparse coefficient with $x\in\mathbb{R}^d$.

For a matrix $D$, a tight frame is formed by its $d$ columns $D_1,\cdots,D_d$, namely,
\begin{align}
\notag \sum^d_{i=1}D_i\left<f,D_i\right>=f~\mbox{and}~\sum^d_{i=1}|\left<f,D_i\right>|^2=\|f\|^2_2
\end{align}
for all $f\in\mathbb{R}^N$, where $\left<\cdot,\cdot\right>$ stands for the standard Euclidean inner product. For all $q\in[1,\infty)$, $x\in\mathbb{R}^N$, let $\|x\|_q=\sqrt[q]{\sum^N_{i=1}|x_i|^q}$ and $\|x\|_{\infty}=\bigvee^N_{i=1}|x_i|$ denote the maximum of $\{|x_1|,\cdots,|x_N|\}.$ One can easily prove that for all $f\in\mathbb{R}^N$,
\begin{align}
\notag DD^*=I_n,~\|f\|^2_2=\|D^*f\|^2_2,
\end{align}
where $D^*$ denotes the transpose of the matrix $D$ and $I_n$ is an identity matrix of $n$ order.

Our goal in the present paper is to recover
the true signal $f\in\mathbb{R}^N$ from $M$ linear measurements $b=\Phi f+z$ in the case that the signal is sparse or approximately sparse in terms of $D$. The $l_1$-analysis approach below is utilized for the recovery of such signal
\begin{align}\label{eq.1}
\tilde{f}=\arg\min_{f\in\mathbb{R}^N}\|D^*f\|_1~\mbox{subject~to}~\Phi f-b\in\mathcal{B},
\end{align}
where $\mathcal{B}$ is a bounded set relied on the noise structure. In this paper, we discuss two types of bounded noise: $l_2$ bounded $\mathcal{B}_0(\varepsilon)=\{z:\|z\|_2\leq\varepsilon\}$ and Dantzig selector bounded $\mathcal{B}_1(\zeta)=\{z:\|D^*\Phi^*z\|_{\infty}\leq\zeta\}.$
In order to discuss the performance of the above method, we will present the concept of $D$-RIP of a sensing matrix, which was first proposed by Cand$\acute{e}$s et al. \cite{Candes et al}. In fact, it is a generalization to the standard RIP.

\begin{definition}
(D-RIP) Let $D$ be a matrix with the size $N\times d$ $(N\leq d)$. One says that a measurement matrix $\Phi$ satisfies the restricted isometry property adapted to $D$ ($D$-RIP for short) of $s$ order with constant $\delta_s$ if
\begin{align}\label{eq.2}
\sqrt{1-\delta_s}\|Dx\|_2\leq\|\Phi Dx\|_2\leq\sqrt{1+\delta_s}\|Dx\|_2
\end{align}
holds for all $s$-sparse vector $x$ in $\mathbb{R}^d$. We call a vector $x\in\mathbb{R}^d$ $s$-sparse if $|\sup(x)|\leq s$, where $\sup(x)=\{i:x_i\neq0\}$ represents the support of the vector $x$. Define the $D$-RIP constant $\delta_s$ as the smallest number $\delta_s$ obeying (\ref{eq.2}) for any $s$-sparse vector $x\in\mathbb{R}^d$.
\end{definition}
By the way, we give the definition of $l_0$ based on the notion of support that $\|x\|_0=|\sup(x)|$, that is, $\|x\|_0$ counts the number of non-zero entries for vector $x$. In the case of that $D$ is an identity matrix, i.e., $D=I$, the Definition $1.1$ degenerates to the definition of standard RIP.

It has been shown that sparse signals in terms of $D$ can be recovered though $l_1$-analysis approach under a variety of conditions on the $D$-RIP. Cand$\acute{e}$s et al. \cite{Candes et al} proved that a sufficient condition $\delta_{2s}<0.08$ can guarantee the reconstruction of sparse signals. In \cite{Li and Lin,Lin et al} respectively improved the upper bound to $\delta_{2s}<0.4931$ and $\delta_{s}<0.307$. Liu et al. \cite{Liu et al} employed the assumption $9\delta_{2s}+4\delta_{4s}<5$ to assure recovery under the general frame. Zhang and Li \cite{Zhang and Li 2015} refined the bound to $\delta_{2s}<\sqrt{2}/2\approx0.707$ and $\delta_{s}<1/3\approx0.333$.  Furthermore, Chen and Li \cite{Chen and Li} gave a high order condition on the $D$-RIP for the recovery of signal.

In this paper, we provide a sufficient condition on D-RIP constant $\delta_{ts}$ under which signals from undersampled data that are approximately sparse in terms of $D$ are guaranteed to be stably reconstructed in the noise situation or exactly recovered in the noiseless situation via the $l_1$-analysis method. We prove that under the condition $\delta_{ts}<t/(4-t)$ for $0<t<4/3$, any signals $f$ which are sparse with respect to a tight frame $D$ can be accurately and stably recovered via (\ref{eq.1}). When $t=1$, our main results are consistent with Theorem $3.1$ and Theorem $4.1$ in \cite{Zhang and Li 2015}. Besides, in the situation of $D=I$ (for the standard compressing sensing), we obtain same results as the main results in \cite{Zhang and Li 2017} and the bound on the constant $\delta_{ts}$ is sharp, referred to see \cite{Cai and Zhang 2014}. Observe that our new bound $\delta_{ts}<t/(4-t)~(t\in(0,4/3))$ improves the condition $\delta_{s}<0.307$ given in \cite{Lin et al} in the case when $t$ is equal to $1$. Weakening the $D$-RIP condition brings a few advantages \cite{Lin et al}. First of all, it allows more sensing matrices to be utilized in compressed sensing. Second, it provides better error estimation in a usual issue to reconstruct signals with noise. Last, improving the $D$-RIP condition allows estimating a sparse signal with more nonzero entries.

The remainder of this paper is organized as following. Some key lemmas and notations are presented in section $2$. The main results are establish in section $3$.
 In section $4$, we provide proofs of theorems. Conclusions are given in the section $5$.

\section{Some technical lemmas}
\label{sec2}
Define $D$ as a tight frame with the size $N\times d$ ($N\leq d$). This means that all row vectors are orthonormal. Denote by $D^{\bot}$ its orthonormal complement, then it is also a tight frame. Therefore, for all $x\in\mathbb{R}^d$,
\begin{align}
\notag \sqrt{\|Dx\|^2_2+\|D^{\bot}x\|^2_2}=\|x\|_2.
\end{align}
The above equality implies that the property of a tight frame $D$: $\|Dx\|_2\leq\|x\|_2.$ It is straightforward to check that if the measurement matrix $\Phi$ obeys D-RIP with constant $\delta_s$, then
\begin{align}\label{eq.3}
\sqrt{1-\delta_s}\|x\|_2\leq\sqrt{\|\Phi Dx\|^2_2+\|D^{\bot}x\|^2_2}\leq\sqrt{1+\delta_s}\|x\|_2
\end{align}
holds for all vectors $x\in\mathbb{R}^d$ with $\|x\|_0\leq s$.

Throughout of this article, we utilize the following notations. Set $\tilde{f}=f+h$ be a solution to (\ref{eq.1}), where $f$ is the unknown signal we wish to recover. For $T\subset\{1,2,\cdots,d\}$, $D_T$ indicates the matrix $D$ limited to the columns indexed by $T$ and $T^c$ represents the complement of $T$ in $\{1,2,\cdots,d\}$, i.e., $T^c=\{1,2,\cdots,d\}\verb|\|T.$ Set $D^*_T=(D_T)^*$ and $D^*h=(x_1,x_2,\cdots,x_d)$. Suppose that $\{|x_i|\}^d_{i=1}$ is a non-increasing sequence, rearranging the indices if necessary. Let $T_0=\{1,2,\cdots,s\}$. Let $S_0$ stand for the index set of the largest $s$ components of $D^*f$ in amplitude. Denote by $x_{[s]}$ the vector comprising the $s$ largest absolute-value coefficients of $x\in\mathbb{R}^d$: $x_{[s]}=\arg\min_{\|y\|_0\leq s}\|x-y\|_2$.

The following lemmas are used in the proofs of main results.
\begin{lemma}
 (Lemma $1$ in \cite{Zhang and Li 2017}) Suppose $\Pi$ is an index set with $|\Pi|=s$. Given vectors $\{u_j:~j\in\Pi\}$ are in a vector space $X$ with inner product $\left<,\cdot,\right>$. Select all subset $\Pi_i\subset\Pi$ satisfying $|\Pi_i|=k,i\in U$ with $|U|=C^k_s$ $(C^k_s=(^s_k))$, then
\begin{align}\label{eq.4}
\sum_{i\in U}\sum_{j\in\Pi_i}u_j=C^{k-1}_{s-1}\sum_{i\in\Pi}u_i,
\end{align}
holds for $k\geq1$.
\end{lemma}
In \cite{Cai and Zhang 2014}, Cai and Zhang introduced a crucial technical tool, which expresses points in a polytope by convex combinations of sparse vectors. The proofs of theorems for this paper depend on this new technique which is stated as the following lemma.
\begin{lemma}
Assume that $\alpha$ is a positive number and $s$ is  a positive integer. The polytope $\tau(\alpha,s)\subset\mathbb{R}^d$ is defined by
$$\tau(\alpha,s)=\{x\in\mathbb{R}^d:\|x\|_{\infty}\leq\alpha,\|x\|_1\leq s\alpha\}.$$ The set of sparse vectors $\mathcal{U}(\alpha,s,x)\subset\mathbb{R}^d$ is defined by
$$\mathcal{U}(\alpha,s,x)=\{u\in\mathbb{R}^d:~\sup(u)\subseteq \sup(x),\|u\|_0\leq s,\|u\|_1=\|x\|_1,\|u\|_{\infty}\leq\alpha\}$$ for all $x\in\mathbb{R}^d$.
Then we can represent any $x\in\tau(\alpha,s)$ as
$$x=\sum_i\lambda_iu_i,$$
where $u_i\in\mathcal{U}(\alpha,s,x)$, $0\leq\lambda_i\leq1$, $\sum_i\lambda_i=1$, and $\sum_i\lambda_i\|u_i\|^2_2\leq s\alpha^2.$
\end{lemma}

\begin{lemma}
We get that
\begin{align}\label{eq.6}
\|D^*_{\Pi}h\|_1+2\|D^*_{\Pi^c}f\|_1\geq\|D^*_{\Pi^c}h\|_1
\end{align}
for any subset $\Pi\subset\{1,\cdots,d\}$.
\end{lemma}
\noindent \textbf{Proof}
According to the minimality of $\tilde{f}$, one implies that
\begin{align}
\notag\|D^*f\|_1\geq\|D^*\tilde{f}\|_1.
\end{align}
That is,
\begin{align}
\notag\|D^*_{\Pi}f\|_1+\|D^*_{\Pi^c}f\|_1\geq\|D^*_{\Pi}\tilde{f}\|_1+\|D^*_{\Pi^c}\tilde{f}\|_1.
\end{align}
Applying the inverse triangular inequality to the above inequality, we get
\begin{align}
\notag\|D^*_{\Pi}f\|_1+\|D^*_{\Pi^c}f\|_1\geq\|D^*_{\Pi}f\|_1-\|D^*_{\Pi}h\|_1+\|D^*_{\Pi^c}h\|_1-\|D^*_{\Pi^c}f\|_1,
\end{align}
which deduces the desired result.

\qed

\section{Main results}
\label{sec3}
In the present section, we will establish new $D$-RIP conditions for the stable reconstruction of sparse signals in terms of $D$ via the $l_1$-analysis method.

\noindent \text{ A. Bounded Noise}

In this subsection, we discuss recovery of signals which are sparse with respect to $D$ in the noisy situation. Specially, we think over two types of bounded noise: $l_2$ bound $\mathcal{B}_0(\varepsilon)=\{z:\|z\|_2\leq\varepsilon\}$ and Dantzig selector bound $\mathcal{B}_1(\zeta)=\{z:\|D^*\Phi^*z\|_{\infty}\leq\zeta\}.$ For simplicity, let $\tilde{f}^{ABP}$ represent the solution of (\ref{eq.1}) obeying $\mathcal{B}=\mathcal{B}_0(\varepsilon)$ and $\tilde{f}^{ADS}$ stand for the solution of (\ref{eq.1}) such that $\mathcal{B}=\mathcal{B}_1(\zeta)$.

\begin{theorem}
Suppose that $D$ denotes an arbitrary tight frame. If the measurement matrix $\Phi$ fulfils the $D$-RIP with constant $\delta_{ts}<t/(4-t)$ for  $0<t<4/3$, then
\begin{align}
\notag\|\tilde{f}^{ABP}-f\|_2\leq A\varepsilon+B\frac{\|D^*f-(D^*f)_{[s]}\|_1}{\sqrt{s}},
\end{align}
and
\begin{align}\label{eq.7}
\|\tilde{f}^{ADS}-f\|_2\leq C\zeta+B\frac{\|D^*f-(D^*f)_{[s]}\|_1}{\sqrt{s}},
\end{align}
where
\begin{align}
\notag A&=\frac{2\sqrt{2}\sqrt{1+\delta_{ts}}\tilde{t}}{t+(t-4)\delta_{ts}},\\
\notag B&=\frac{4\sqrt{2}\delta_{ts}
+2\sqrt{2}\sqrt{(t+(t-4)\delta_{ts})\delta_{ts}}}{t+(t-4)\delta_{ts}}+\frac{\sqrt{2}}{2},
\end{align}
and
$$C=\frac{2\sqrt{2s}\tilde{t}}{t+(t-4)\delta_{ts}}.$$
\end{theorem}

\begin{remark}
When $D$ is an identity matrix, namely for the standard compressed sensing, we obtain same results as Theorem $2$ in \cite{Zhang and Li 2017} and the condition is the weakest, for more information, see \cite{Cai and Zhang 2014}.
\end{remark}

\begin{remark}
In the case when $t$ is equal to $1$, Theorem $3.1$ coincides with Theorem $4.1$ in \cite{Zhang and Li 2015}. The sufficient condition  $\delta_{ts}<t/(4-t)~(t\in(0,4/3))$ extends the condition $\delta_s<1/3$ to a more general context. Furthermore, the error estimations (\ref{eq.7}) is much better than $(4.1)$ and $(4.2)$ of Theorem $4.1$ in \cite{Zhang and Li 2015}.
\end{remark}

\begin{remark}
It is known that Lin et al. \cite{Lin et al} gave a sufficient condition $\delta_{s}<0.307$, which guarantees that $s$-sparse signals in term of $D$ can be stably reconstructed. When $t=1$, the result of Theorem $3.1$ improves the bound to $\delta_{s}<1/3$, for more details, see Remark $III.2.(d)$ in \cite{Lin et al}.
\end{remark}

\begin{remark}
In the proof of Theorem $3.1$, by applying Lemma $5.3$ in \cite{Cai and Zhang 2013} to the estimation of $\|D^*_{T^c}h\|_2$, we obtain different error estimations as follows:
\begin{align}
\notag\|\tilde{f}^{ABP}-f\|_2\leq A\varepsilon+\tilde{B}\frac{\|D^*f-(D^*f)_{[s]}\|_1}{\sqrt{s}},
\end{align}
and
\begin{align}\label{eq.34}
\|\tilde{f}^{ADS}-f\|_2\leq C\zeta+\tilde{B}\frac{\|D^*f-(D^*f)_{[s]}\|_1}{\sqrt{s}},
\end{align}
where
\begin{align}
\notag \tilde{B}&=\frac{4\sqrt{2}\delta_{ts}
+2\sqrt{2}\sqrt{(t+(t-4)\delta_{ts})\delta_{ts}}}{t+(t-4)\delta_{ts}}+2,
\end{align}
and $A,~C$ are defined by Theorem $3.1.$ Obviously, note that the upper bounds of error estimations determined by (\ref{eq.7}) is much better than those presented by (\ref{eq.34}). In addition, (\ref{eq.34}) is the same as Theorem $4.1$ in \cite{Zhang and Li 2015} in the situation that $t$ is equal to $1$.
\end{remark}

In the following corollary, we consider exact reconstruction of sparse (in terms of $D$) signals with noiseless via the approach (\ref{eq.1}) with $\mathcal{B}=\{0\}$.

\begin{corollary}
Assume that one observes $b=\Phi f$ with $\|D^*f\|_0\leq s$. The solution to (\ref{eq.1}) satisfying $\mathcal{B}=\{0\}$ estimates the unknown signal $f$, viz, $\tilde{f}=f$ provided that the matrix $\Phi$ meets the $D$-RIP with constant $\delta_{ts}<t/(4-t)$ for $0<t<4/3$.
\end{corollary}

\begin{remark}
In the noise-free situation when noise vector $z=0$ and the unknown signal $f$ is $s$-sparse with respect to $D$, the result directly follows from Theorem $3.1.$
\end{remark}

\begin{remark}
When $D$ is an identity matrix, the result is the same as Theorem $1$ in \cite{Zhang and Li 2017}.
\end{remark}

\noindent \text{ B. Gaussian Noise}

In statistics and signal processing, it is of interest to study the signal reconstruction with the error vector obeying Gaussian noise (i.e., $z\sim\mathbb{N}(0,\sigma^2I)$). Define two bounded sets
\begin{align}
\notag\mathcal{B}_2=\left\{z:\|z\|^2_2\leq\sigma^2\left(M+2\sqrt{M\log M}\right)\right\}
\end{align}
and
\begin{align}
\notag\mathcal{B}_3=\left\{z:\|D^*\Phi^*z\|^2_{\infty}\leq8\sigma^2\log d\right\}.
\end{align}
The following lemma shows that the Gaussian error is bounded.
\begin{lemma} (Lemma $III.3$ in \cite{Lin et al})
Suppose a matrix $\Phi\in\mathbb{R}^{M\times N}$ fulfils the $D$-RIP with real number $\delta_1\in(0,1)$, then Gaussian noise $z$ obeys
\begin{align}
\notag1-\mathcal{P}(z\in\mathcal{B}_2)\leq\frac{1}{M}
\end{align}
and
\begin{align}
\notag1-\mathcal{P}(z\in\mathcal{B}_3)\leq\frac{1}{d\sqrt{2\pi\log d}},
\end{align}
where the noise vector $z\sim\mathbb{N}(0,\sigma^2I)$.
\end{lemma}
\begin{theorem}
We suppose that $D$ is an arbitrary tight frame. Let $\tilde{f}_1$ represent the solution of (\ref{eq.1}) obeying $\mathcal{B}=\mathcal{B}_2$ and $\tilde{f}_2$ stand for the solution of (\ref{eq.1}) such that $\mathcal{B}=\mathcal{B}_3$. If the measurement matrix $\Phi$ fulfils the $D$-RIP with constant $\delta_{ts}<t/(4-t)$ for $0<t<4/3$, then
\begin{align}
\notag\|\tilde{f}_1-f\|_2\leq A\sigma\sqrt{M+2\sqrt{M\log M}}+B\frac{\|D^*f-(D^*f)_{[s]}\|_1}{\sqrt{s}}
\end{align}
with probability not less than $1-1/M$ and
\begin{align}
\notag\|\tilde{f}_2-f\|_2\leq 2\sqrt{2}C\sigma\sqrt{\log d}+B\frac{\|D^*f-(D^*f)_{[s]}\|_1}{\sqrt{s}}
\end{align}
with probability greater or equal to $1-1/(d\sqrt{2\pi\log d})$. Here, the constants $A,~B$ and $C$ are defined by Theorem $3.1$.
\end{theorem}
\begin{remark}
Since in this situation noise vector $z$ is in some bounded set with high probability, the theorem follows from Theorem $3.1$ and Lemma $3.1$.
\end{remark}

\begin{remark}
In the situation when $D$ is equal to an identity matrix $I$, Theorem $3.2$ is the same as Corollary $2$ in \cite{Zhang and Li 2017}.
\end{remark}

\section{Proofs of main results}
\label{sec4}

\noindent \textbf{Proof of Theorem $3.1$.}

We suppose that $ts$ is an integer. Observe that $\|D^*_{S_0}h\|_1\leq\|D^*_{T_0}h\|_1$, $\|D^*_{T^c_0}h\|_1\leq\|D^*_{S^c_0}h\|_1$. It thus follows from Lemma $2.3$ that
\begin{align}\label{eq.9}
\|D^*_{T_0}h\|_1+2\|D^*_{S^c_0}f\|_1\geq\|D^*_{T^c_0}h\|_1.
\end{align}
Set
$$\frac{\|D^*_{T_0}h\|_1+2\|D^*_{S^c_0}f\|_1}{s}=r.$$
Then
$$r^2=\frac{\|D^*_{T_0}h\|^2_1+4\|D^*_{T_0}h\|_1\|D^*_{S^c_0}f\|_1+4\|D^*_{S^c_0}f\|^2_1}{s^2}.$$
By employing $\|D^*_{T_0}h\|^2_1\leq s\|D^*_{T_0}h\|^2_2$ to the above equality, we get
\begin{align}\label{eq.10}
r^2\leq\frac{\|D^*_{T_0}h\|^2_2}{s}+\frac{4\|D^*_{T_0}h\|_2\|D^*_{S^c_0}f\|_1}{s^{\frac{3}{2}}}+\frac{4\|D^*_{S^c_0}f\|^2_1}{s^2}.
\end{align}
Pick out positive integers $m$ and $n$ contenting $n\leq m\leq s$ and $(m+n)/s=t$. We use subsets $T_i,S_j\subset\{1,\cdots,s\}$ to represent all probable index sets with $|T_i|=m~(i\in U)$, $|S_j|=n~(j\in V)$ that $|U|=C^m_s$ and $|V|=C^n_s$.

From the definition of $T_0$, we have
\begin{align}\label{eq.11}
\|D^*_{T^c_0}h\|_{\infty}\leq\frac{\|D^*_{T_0}h\|_1}{s}\leq\frac{\|D^*_{T_0}h\|_1+2\|D^*_{S^c_0}f\|_1}{s}\leq\frac{s}{n}r.
\end{align}
Moreover, it is known that the $l_1$ norm of $D^*_{T^c_0}h$ is bounded, i.e.,
$$\|D^*_{T^c_0}h\|_1\leq n\frac{s}{n}r.$$
 Hence, by Lemma $2.2$, we imply that
\begin{align}
\notag D^*_{T^c_0}h=\sum_i\lambda_iu_i,
\end{align}
where $u_i$ is $n$-sparse, i.e., $$\|u_i\|_0=|\sup(u_i)|\leq n,$$ and $\sum_i\lambda_i=1$ with for each $i$, $\lambda_i\in(0,1]$, $$\sup(u_i)\subset\sup(D^*_{T^c_0}h),~\|u_i\|_1=\|D^*_{T^c_0}h\|_1,~\|u_i\|_{\infty}\leq\frac{s}{n}r,$$
\begin{align}\label{eq.12}
\sum_i\lambda_i\|u_i\|^2_2\leq n\left(\frac{s}{n}r\right)^2=\frac{s^2r^2}{n}.
\end{align}
Analogously, we can also decompose $D^*_{T^c_0}h$ into
\begin{align}
\notag D^*_{T^c_0}h=\sum_i\mu_iv_i,
\end{align}
and
\begin{align}
\notag D^*_{T^c_0}h=\sum_i\nu_iw_i.
\end{align}
Here $|\sup(v_i)|\leq m$, $|\sup(w_i)|\leq (t-1)s~(t>1)$ and
\begin{align}\label{eq.13}
&\sum_i\mu_i\|v_i\|^2_2\leq\frac{s^2r^2}{m},\\
\label{eq.14}
&\sum_i\nu_i\|w_i\|^2_2\leq\frac{sr^2}{t-1}.
\end{align}
 By the feasibility of $\tilde{f}^{ABP}$, we get that
\begin{align}\label{eq.15}
\|\Phi h\|_2\leq\|\Phi (\tilde{f}^{ABP}-f)\|_2\leq\|\Phi\tilde{f}^{ABP}-b\|_2+\|\Phi f-b\|_2\leq2\varepsilon.
\end{align}
Thus, it follows that
\begin{align}
\notag|\left<\Phi DD^*_{T_0}h,\Phi h\right>|^2&\leq\|\Phi DD^*_{T_0}h\|^2_2\|\Phi h\|^2_2\\
\notag&\leq4\varepsilon^2(1+\delta_s)\|DD^*_{T_0}h\|^2_2\\
\label{eq.16}&\leq4\varepsilon^2(1+\delta_{ts})\|D^*_{T_0}h\|^2_2
\end{align}
for $t\in[1,4/3)$, where for the third inequality, we used the fact that $\delta_s\leq\delta_{s_1}$ if $s\leq s_1$\cite{Lin et al};
\begin{align}
\notag|\left<\Phi DD^*_{T_0}h,\Phi h\right>|^2&\leq4\varepsilon^2(1+\delta_s)\|DD^*_{T_0}h\|^2_2\\
\notag&\leq4\varepsilon^2\left(1+\left(\frac{2}{t}-1\right)\delta_{ts}\right)\|D^*_{T_0}h\|^2_2\\
\label{eq.17}&\leq\frac{4\varepsilon^2(1+\delta_{ts})}{t}\|D^*_{T_0}h\|^2_2
\end{align}
for $t\in(0,1)$, where for the second inequality follows from Lemma $4.1$ in \cite{Cai and Zhang 2013}.

For notational simplicity, set
\begin{align}
\notag \Delta_0=&\sum_{i\in U,k}\frac{s-n}{mC^m_s}\lambda_k\left(m^2\|E(D^*_{T_i}h+\frac{n}{s}u_k)\|^2_2-n^2\|E(D^*_{T_i}h-\frac{m}{s}u_k)\|^2_2\right)\\
\label{eq.18}&+\sum_{j\in V,k}\frac{s-m}{nC^n_s}\mu_k\left(n^2\|E(D^*_{S_j}h+\frac{m}{s}v_k)\|^2_2-m^2\|E(D^*_{S_j}h-\frac{n}{s}v_k)\|^2_2\right),
\end{align}
where $E\in \mathbb{R}^{L\times d}$ and $L$ is an arbitrary integer. Let $\rho(m,n)=(m-n)^2+2(t-2)mn$.
The construction of following two identities make use of ideals from equalities $(14)$ and $(15)$ in \cite{Zhang and Li 2017}. One can verify that
\begin{align}
\notag \frac{\rho(m,n)(t-1)}{mnC^m_sC^n_{s-m}}&\sum_{T_i\bigcap S_j=\phi}\left(\frac{mn}{t-1}\|E(D^*_{T_i}h+D^*_{S_j}h)\|^2_2+\|E(nD^*_{T_i}h-mD^*_{S_j}h)\|^2_2\right)\\
\notag&=t\Delta_0+2mn(t-2)t^2\left<ED^*_{T_0}h,ED^*h\right>
\end{align}
holds for $t\in(0,1),$ and
\begin{align}
\notag \rho(m,n)&\sum_k\nu_k\bigg\{\|E(D^*_{T_0}h+(t-1)w_k)\|^2_2-\|(t-1)E(D^*_{T_0}h-w_k)\|^2_2\bigg\}\\
\notag&=-(3t-4)\Delta_0+2\{(t-1)s^2-mn\}t^3\left<E D^*_{T_0}h,E D^*h\right>
\end{align}
holds for $t\in[1,4/3).$ It thus follows that
\begin{align}
\notag \frac{\rho(m,n)(t-1)}{mnC^m_sC^n_{s-m}}\sum_{T_i\bigcap S_j=\phi}&\bigg(\frac{mn}{t-1}\|\Phi D(D^*_{T_i}h+D^*_{S_j}h)\|^2_2+\|\Phi D(nD^*_{T_i}h-mD^*_{S_j}h)\|^2_2\\
\notag&+\frac{mn}{t-1}\|D^{\bot}(D^*_{T_i}h+D^*_{S_j}h)\|^2_2+\|D^{\bot}(nD^*_{T_i}h-mD^*_{S_j}h)\|^2_2\bigg)\\
\label{eq.19}&=t\tilde{\Delta}_0+2mn(t-2)t^2\left<\Phi DD^*_{T_0}h,\Phi h\right>
\end{align}
holds for $t\in(0,1),$ and
\begin{align}
\notag \rho(m,n)\sum_k\nu_k&\bigg(\|\Phi D(D^*_{T_0}h+(t-1)w_k)\|^2_2-\|(t-1)\Phi D(D^*_{T_0}h-w_k)\|^2_2\\
\notag&+\|D^{\bot}(D^*_{T_0}h+(t-1)w_k)\|^2_2-\|(t-1)D^{\bot}(D^*_{T_0}h-w_k)\|^2_2\bigg)\\
\label{eq.20}&=-(3t-4)\tilde{\Delta}_0+2\{(t-1)s^2-mn\}t^3\left<\Phi DD^*_{T_0}h,\Phi h\right>
\end{align}
holds for $t\in[1,4/3)$, where
\begin{align}
\notag \tilde{\Delta}_0=&\sum_{i\in U,k}\frac{s-n}{mC^m_s}\lambda_k\bigg(m^2\|\Phi D(D^*_{T_i}h+\frac{n}{s}u_k)\|^2_2-n^2\|\Phi D(D^*_{T_i}h-\frac{m}{s}u_k)\|^2_2\\
\notag&+m^2\|D^{\bot}(D^*_{T_i}h+\frac{n}{s}u_k)\|^2_2-n^2\|D^{\bot}(D^*_{T_i}h-\frac{m}{s}u_k)\|^2_2\bigg)\\
\notag&+\sum_{j\in V,k}\frac{s-m}{nC^n_s}\mu_k\bigg(n^2\|\Phi D(D^*_{S_j}h+\frac{m}{s}v_k)\|^2_2-m^2\|\Phi D(D^*_{S_j}h-\frac{n}{s}v_k)\|^2_2\\
\label{eq.21}&+n^2\|D^{\bot}(D^*_{S_j}h+\frac{m}{s}v_k)\|^2_2-m^2\|D^{\bot}(D^*_{S_j}h-\frac{n}{s}v_k)\|^2_2\bigg),
\end{align}
and we used the fact that $\left<D^{\bot}D^*_{T_0}h,D^{\bot}D^*h\right>=0$. For $\rho(m,n)$, if $ts$ is odd, then let $m=n+1=(ts+1)/2$; if $ts$ is even, set $m=n=ts/2$. Easily check that $\rho(m,n)<0$ for both situations.

According to $\|D^*_{T_i}h\|_0=\|v_k\|_0\leq m$, $\|D^*_{S_j}h\|_0=\|u_k\|_0\leq n$ ($m+n=ts$) and combining with (\ref{eq.3}), we imply that
\begin{align}
\notag \tilde{\Delta}_0\geq&\sum_{i\in
 U,k}\frac{s-n}{mC^m_s}\lambda_k\bigg(m^2(1-\delta_{ts})\|D^*_{T_i}h+\frac{n}{s}u_k\|^2_2
-n^2(1+\delta_{ts})\|D^*_{T_i}h-\frac{m}{s}u_k\|^2_2\bigg)\\
\notag&+\sum_{j\in V,k}\frac{s-m}{nC^n_s}\mu_k\bigg(n^2(1-\delta_{ts})\|D^*_{S_j}h+\frac{m}{s}v_k\|^2_2-m^2(1+\delta_{ts})\|D^*_{S_j}h-\frac{n}{s}v_k\|^2_2\bigg).
\end{align}
Notice that $\left<D^*_{T_i}h,u_k\right>=\left<D^*_{S_j}h,v_k\right>=0$ for each $i,j,k$. Then
\begin{align}
\notag \tilde{\Delta}_0\geq&\sum_{i\in U,k}\frac{s-n}{mC^m_s}\lambda_k\bigg(m^2(1-\delta_{ts})\|D^*_{T_i}h\|^2_2
+\frac{m^2n^2}{s^2}(1-\delta_{ts})\|u_k\|^2_2-n^2(1+\delta_{ts})\|D^*_{T_i}h\|^2_2\\
\notag&-\frac{m^2n^2}{s^2}(1+\delta_{ts})\|u_k\|^2_2\bigg)\\
\notag&+\sum_{j\in V,k}\frac{s-m}{nC^n_s}\mu_k\bigg(n^2(1-\delta_{ts})\|D^*_{S_j}h\|^2_2
+\frac{m^2n^2}{s^2}(1-\delta_{ts})\|v_k\|^2_2-m^2(1+\delta_{ts})\|D^*_{S_j}h\|^2_2\\
\notag&-\frac{m^2n^2}{s^2}(1+\delta_{ts})\|v_k\|^2_2\bigg)\\
\notag&=\frac{s-n}{mC^m_s}\{(m+n)(m-n)-(m^2+n^2)\delta_{ts})\}\sum_{i\in U}\|D^*_{T_i}h\|^2_2
-\frac{2(s-n)mn^2\delta_{ts}}{s^2}\sum_{k}\lambda_k\|u_k\|^2_2\\
\notag&+\frac{s-m}{nC^n_s}\{-(m+n)(m-n)-(m^2+n^2)\delta_{ts}\}\sum_{j\in V}\|D^*_{S_j}h\|^2_2
-\frac{2(s-m)m^2n\delta_{ts}}{s^2}\sum_{k}\mu_k\|v_k\|^2_2.
\end{align}
By exploiting (\ref{eq.4}) to the above inequality, we get
\begin{align}
\notag \tilde{\Delta}_0\geq&\{(m+n)(m-n)-(m^2+n^2)\delta_{ts}\}\frac{s-n}{mC^m_s}C^{m-1}_{s-1}\|D^*_{T_0}h\|^2_2-\frac{2(s-n)mn^2\delta_{ts}}{s^2}\sum_{k}\lambda_k\|u_k\|^2_2\\
\notag&+\{-(m+n)(m-n)-(m^2+n^2)\delta_{ts}\}\frac{s-m}{nC^n_s}C^{n-1}_{s-1}\|D^*_{T_0}h\|^2_2-\frac{2(s-m)m^2n\delta_{ts}}{s^2}\sum_{k}\mu_k\|v_k\|^2_2.
\end{align}
By utilizing (\ref{eq.12}) and (\ref{eq.13}) to the above inequality, we get that
\begin{align}
\notag \tilde{\Delta}_0\geq&\{(m+n)(m-n)-(m^2+n^2)\delta_{ts}\}\frac{s-n}{s}\|D^*_{T_0}h\|^2_2-\frac{2(s-n)mn^2\delta_{ts}}{s^2}\frac{s^2r^2}{n}\\
\notag&+\{-(m+n)(m-n)-(m^2+n^2)\delta_{ts}\}\frac{s-m}{s}\|D^*_{T_0}h\|^2_2-\frac{2(s-m)m^2n\delta_{ts}}{s^2}\frac{s^2r^2}{m}\\
\notag=&\bigg(\frac{(m+n)(m-n)^2}{s}-\frac{(m^2+n^2)(2s-m-n)\delta_{ts}}{s}\bigg)\|D^*_{T_0}h\|^2_2\\
\notag&-2mn(2s-m-n)\delta_{ts}r^2\\
\label{eq.22}=&\{(m-n)^2t+(m^2+n^2)(t-2)\delta_{ts}\}\|D^*_{T_0}h\|^2_2+2mns(t-2)\delta_{ts}r^2.
\end{align}
For convenience, denote
\begin{align}
\notag \mathcal{F}&=\notag \frac{\rho(m,n)(t-1)}{mnC^m_sC^n_{s-m}}\sum_{T_i\bigcap S_j=\phi}\bigg(\frac{mn}{t-1}\|\Phi D(D^*_{T_i}h+D^*_{S_j}h)\|^2_2+\|\Phi D(nD^*_{T_i}h-mD^*_{S_j}h)\|^2_2\\
\notag&+\frac{mn}{t-1}\|D^{\bot}(D^*_{T_i}h+D^*_{S_j}h)\|^2_2+\|D^{\bot}(nD^*_{T_i}h-mD^*_{S_j}h)\|^2_2\bigg),\\
\notag \mathcal{G}&=t\tilde{\Delta}_0+2mn(t-2)t^2\left<\Phi DD^*_{T_0}h,\Phi h\right>,\\
\notag \mathcal{H}&=\rho(m,n)\sum_k\nu_k\bigg(\|\Phi D(D^*_{T_0}h+(t-1)w_k)\|^2_2-\|(t-1)\Phi D(D^*_{T_0}h-w_k)\|^2_2\\
\notag&+\|D^{\bot}(D^*_{T_0}h+(t-1)w_k)\|^2_2-\|(t-1)D^{\bot}(D^*_{T_0}h-w_k)\|^2_2\bigg),\\
\notag \mathcal{I}&=-(3t-4)\tilde{\Delta}_0+2\{(t-1)s^2-mn\}t^3\left<\Phi DD^*_{T_0}h,\Phi h\right>,
\end{align}
where $\tilde{\Delta}_0$ is determined by (\ref{eq.21}).

We firstly take into the situation of $t\in(0,1)$.

Since $D^*_{T_i}h+D^*_{S_j}h$ are $ts$-sparse for each $i,j$, we make use of (\ref{eq.3}), then it gives
\begin{align}
\notag \mathcal{F}&\leq\frac{\rho(m,n)(t-1)}{mnC^m_sC^n_{s-m}}\sum_{T_i\bigcap S_j=\phi}\left(\frac{mn}{t-1}(1-\delta_{ts})\|D^*_{T_i}h+D^*_{S_j}h\|^2_2+(1+\delta_{ts})\|nD^*_{T_i}h-mD^*_{S_j}h\|^2_2\right)\\
\notag &=\frac{\rho(m,n)(t-1)}{mnC^m_sC^n_{s-m}}\bigg\{
\frac{mn}{t-1}(1-\delta_{ts})\bigg(C^n_{s-m}\sum_{i\in U}\|D^*_{T_i}h\|^2_2+C^m_{s-n}\sum_{j\in V}\|D^*_{S_j}h\|^2_2\bigg)\\
\notag&+(1+\delta_{ts})\bigg( n^2C^n_{s-m}\sum_{i\in U}\|D^*_{T_i}h\|^2_2+m^2C^m_{s-n}\sum_{j\in V}\|D^*_{S_j}h\|^2_2\bigg)\bigg\}.
\end{align}
Applying Lemma $2.2$ to the above inequality, we get that
\begin{align}
\notag \mathcal{F}&\leq\frac{\rho(m,n)(t-1)}{mnC^m_sC^n_{s-m}}\bigg\{
\frac{mn}{t-1}(1-\delta_{ts})\bigg( C^n_{s-m}C^{m-1}_{s-1}\|D^*_{T_0}h\|^2_2+C^m_{s-n}C^{n-1}_{s-1}\|D^*_{T_0}h\|^2_2\bigg)\\
\notag&+(1+\delta_{ts})\bigg( n^2C^n_{s-m}C^{m-1}_{s-1}\|D^*_{T_0}h\|^2_2+m^2C^m_{s-n}C^{n-1}_{s-1}\|D^*_{T_0}h\|^2_2\bigg)\bigg\}\\
\label{eq.23}&=\rho(m,n)\{t+(t-2)\delta_{ts}\}t\|D^*_{T_0}h\|^2_2.
\end{align}
A combination of (\ref{eq.17}) and (\ref{eq.22}), we obtain
\begin{align}
\notag\mathcal{G}&\geq \{t(m-n)^2+(m^2+n^2)(t-2)\delta_{ts}\}t\|D^*_{T_0}h\|^2_2+2mnst(t-2)\delta_{ts}r^2\\
\label{eq.24}&+4\varepsilon mn(t-2)t\sqrt{(1+\delta_{ts})t}\|D^*_{T_0}h\|_2.
\end{align}
Combine the above two inequalities (\ref{eq.23}) and (\ref{eq.24}), which leads to
\begin{align}
\notag-2mnst(t-2)\delta_{ts}r^2\geq2mn(t-2)t[-t+(3-t)\delta_{ts}]\|D^*_{T_0}h\|^2_2+4\varepsilon mn(t-2)t\sqrt{(1+\delta_{ts})t}\|D^*_{T_0}h\|_2.
\end{align}
Plug (\ref{eq.10}) into the above inequality, we get
\begin{align}
\notag&2mnt(t-2)\bigg((4-t)\left[\frac{t}{4-t}-\delta_{ts}\right]\|D^*_{T_0}h\|^2_2\\
\label{eq.25}&-\bigg(\frac{4\delta_{ts}\|D^*_{S^c_0}f\|_1}{\sqrt{s}}+2\varepsilon \sqrt{(1+\delta_{ts})t}\bigg)\|D^*_{T_0}h\|_2
-\frac{4\delta_{ts}\|D^*_{S^c_0}f\|^2_1}{s}\bigg)\geq0,
\end{align}
which is a second-order inequality for $\|D^*_{T_0}h\|_2$ with the coefficient of square term is less than zero under the condition $\delta_{ts}<t/(4-t)$.

We now consider the situation of $t\in[1,4/3)$.

Combining with $\rho(m,n)<0$, the definition of the $D$-RIP of $ts$ order and (\ref{eq.3}), we imply
\begin{align}
\notag \mathcal{H}&\leq\rho(m,n)\sum_k\nu_k\bigg((1-\delta_{ts})\|D^*_{T_0}h+(t-1)w_k\|^2_2-(1+\delta_{ts})\|(t-1)(D^*_{T_0}h-w_k)\|^2_2\bigg).
\end{align}
Since the support of $D^*_{T_0}h$ does not intersect with that of $w_k$, one gains
\begin{align}
\notag \mathcal{H} &\leq\rho(m,n)\sum_k\nu_k\bigg((1-\delta_{ts})\|D^*_{T_0}h\|^2_2+(1-\delta_{ts})(t-1)^2\|w_k\|^2_2\\
\notag &-(t-1)^2(1+\delta_{ts})\|D^*_{T_0}h\|^2_2-(t-1)^2(1+\delta_{ts})\|w_k\|^2_2\bigg)\\
\notag &=\rho(m,n)\sum_k\nu_k\bigg([1-(t-1)^2-(1+(t-1)^2)\delta_{ts}]\|D^*_{T_0}h\|^2_2\\
\notag &-2(t-1)^2\delta_{ts}\|w_k\|^2_2\bigg).
\end{align}
Substituting (\ref{eq.14}) into the above inequality, we get that
\begin{align}
\label{eq.26}\mathcal{H}&\leq\rho(m,n)
\bigg\{[1-(t-1)^2-(1+(t-1)^2)\delta_{ts}]\|D^*_{T_0}h\|^2_2-2(t-1)s\delta_{ts}r^2\bigg\}.
\end{align}
Under the condition of Theorem $3.1$, we can check that for $t\in[1,4/3),$
\begin{align}
\label{eq.27}(t-1)s^2<\frac{(2-t)^2s^2-1}{4}+(t-1)s^2=\frac{t^2s^2-1}{4}\leq mn.
\end{align}
By (\ref{eq.16}), (\ref{eq.22}) and (\ref{eq.27}), we derive
\begin{align}
\notag\mathcal{I}&\geq-(3t-4)\bigg([t(m-n)^2+(m^2+n^2)(t-2)\delta_{ts}]\|D^*_{T_0}h\|^2_2+2mns(t-2)\delta_{ts}r^2\bigg)\\
\label{eq.28}&+4\varepsilon t^3\sqrt{1+\delta_{ts}}((t-1)s^2-mn)\|D^*_{T_0}h\|_2.
\end{align}
Combining with (\ref{eq.26}) and (\ref{eq.28}), one gets
\begin{align}
\notag&2t^2s\left[-(t-1)s^2+mn\right]\delta_{ts}r^2\\
\notag&\geq\left\{-2s^2t^3(t-1)+2mnt^2+2t^2(t-3)[-(t-1)s^2+mn]\delta_{ts}\right\}\|D^*_{T_0}h\|^2_2\\
\notag&+4\varepsilon t^3\sqrt{1+\delta_{ts}}[(t-1)s^2-mn]\|D^*_{T_0}h\|_2.
\end{align}
By applying (\ref{eq.10}) to the above inequality, we have
\begin{align}
\notag&2[(t-1)s^2-mn]t^2\bigg\{(4-t)\left(\frac{t}{4-t}-\delta_{ts}\right)\|D^*_{T_0}h\|^2_2\\
\label{eq.29}&-\bigg(\frac{4\delta_{ts}}{\sqrt{s}}\|D^*_{S^c_0}f\|_1+2\varepsilon t\sqrt{1+\delta_{ts}}\bigg)\|D^*_{T_0}h\|_2
-\frac{4\delta_{ts}}{s}\|D^*_{S^c_0}f\|^2_1\bigg\}\geq0,
\end{align}
which is a second-order inequality for $\|D^*_{T_0}h\|_2$. Combining with (\ref{eq.25}) and (\ref{eq.29}), we get
\begin{align}
\notag\|D^*_{T_0}h\|_2\leq&\left(\frac{4\delta_{ts}\|D^*_{S^c_0}f\|_1}{\sqrt{s}}+2\varepsilon \sqrt{1+\delta_{ts}}\tilde{t}\right)\frac{1}{2(t+(t-4)\delta_{ts})}\\
\notag&+\left\{\bigg(\frac{4\delta_{ts}\|D^*_{S^c_0}f\|_1}{\sqrt{s}}+2\varepsilon \sqrt{1+\delta_{ts}}\tilde{t}\bigg)^2
+\frac{16(t+(t-4)\delta_{ts})\delta_{ts}\|D^*_{S^c_0}f\|^2_1}{s}\right\}^{\frac{1}{2}}\\
\notag&\times\frac{1}{2(t+(t-4)\delta_{ts})},
\end{align}
where $\tilde{t}=\max\{t,\sqrt{t}\}$ denotes the maximum of $\{t,\sqrt{t}\}.$ By applying the fact that $(x^2+y^2)^{1/2}\leq x+y$ for $x,y\geq0$ to the above inequality, we get
\begin{align}
\label{eq.30}\|D^*_{T_0}h\|_2\leq\frac{2\varepsilon\sqrt{1+\delta_{ts}}\tilde{t}}{t+(t-4)\delta_{ts}}+\frac{2\left(2\delta_{ts}
+\sqrt{(t+(t-4)\delta_{ts})\delta_{ts}}\right)\|D^*_{S^c_0}f\|_1}{\sqrt{s}(t+(t-4)\delta_{ts})}.
\end{align}
By the common inequality
$$\sum^N_{j=1}|v_j|^2\leq\max_{1\leq j\leq N}|v_j|\sum^N_{j=1}|v_j|,$$
we get
\begin{align}
\notag\|D^*_{T^c_0}h\|_2\leq\sqrt{\|D^*_{T^c_0}h\|_{\infty}\|D^*_{T^c_0}h\|_1}.
\end{align}
By applying $\|D^*_{T^c_0}h\|_{\infty}\leq\|D^*_{T_0}h\|_1/s$ and (\ref{eq.9}) to the above inequality, we get
\begin{align}
\notag\|D^*_{T^c_0}h\|_2\leq\sqrt{\frac{\|D^*_{T_0}h\|^2_1}{s}+\frac{2\|D^*_{T_0}h\|_1\|D^*_{S^c_0}f\|_1}{s}}\\
\label{eq.31}\leq\sqrt{\|D^*_{T_0}h\|^2_2+\frac{2\|D^*_{T_0}h\|_2\|D^*_{S^c_0}f\|_1}{\sqrt{s}}},
\end{align}
where the last inequality, we have utilized $\|D^*_{T_0}h\|_1\leq\sqrt{s}\|D^*_{T_0}h\|_2$.

Finally, a combination of (\ref{eq.30}) and (\ref{eq.31}), we obtain
\begin{align}
\notag\|h\|^2_2&=\|D^*h\|^2_2=\|D^*_{T_0}h\|^2_2+\|D^*_{T^c_0}h\|^2_2\\
\notag&\leq2\|D^*_{T_0}h\|^2_2+\frac{2\|D^*_{T_0}h\|_2\|D^*_{S^c_0}f\|_1}{\sqrt{s}}\\
\notag&\leq\left\{\sqrt{2}\|D^*_{T_0}h\|_2+\frac{\|D^*_{S^c_0}f\|_1}{\sqrt{2s}}\right\}^2\\
\notag&\leq\left\{\frac{2\sqrt{2}\varepsilon\sqrt{1+\delta_{ts}}\tilde{t}}{t+(t-4)\delta_{ts}}+\left(\frac{4\sqrt{2}\delta_{ts}
+2\sqrt{2}\sqrt{(t+(t-4)\delta_{ts})\delta_{ts}}}{(t+(t-4)\delta_{ts})}+\frac{1}{\sqrt{2}}\right)\frac{\|D^*_{S^c_0}f\|_1}{\sqrt{s}}\right\}^2.
\end{align}

If $ts$ is not an integer, denote $\hat{t}s=[ts],$ then $\hat{t}s$ is an integer obeying $\hat{t}>t.$  For $\hat{t}\in(0,4/3)$, we have $\delta_{\hat{t}s}=\delta_{ts}<t/(4-t)<\hat{t}/(4-\hat{t})$. Similar to the above proof, one can prove the result by dealing with $\delta_{\hat{t}s}$.

For the situation of Dantzig selector bounded noise:

According to the feasibility of $\tilde{f}^{ADS}$:
\begin{align}
\notag\|D^*\Phi^*\Phi h\|_{\infty}&\leq\|D^*\Phi^*(b-\Phi\tilde{f}^{ADS})\|_{\infty}+\|D^*\Phi^*(\Phi f-b)\|_{\infty}\\
\label{eq.32}&\leq2\zeta
\end{align}
It consequently follows that
\begin{align}
\notag|\left<\Phi h,\Phi DD^*_{T_0}h\right>|^2&\leq|\left<D^*\Phi^*\Phi h,D^*_{T_0}h\right>|^2\\
\notag&\leq\|D^*\Phi^*\Phi h\|^2_{\infty}\|D^*_{T_0}h\|^2_1\\
\label{eq.33}&\leq4s\zeta^2\|D^*_{T_0}h\|^2_2.
\end{align}
The rest of proof is similar to the $l_2$ bounded noise situation. The proof of Theorem $3.1$ is complete.

\qed

\section{Conclusions}
This paper considers a sufficient condition concerning the restricted isometry property adapted to a tight frame $D$ for the reconstruction of signals. We show that the $D$-RIP constant $\delta_{ts}$ from the measurement matrix $\Phi$ obeys $\delta_{ts}<t/(4-t)$ for $t\in(0,4/3)$, signals that are nearly $s$-sparse with respect to $D$ can be stably estimated by $l_1$-analysis methods. When $t=1$, our main results coincide with Theorems $3.1,~4.1$ in \cite{Zhang and Li 2015} and the error estimates are smaller than those given in their work. Moreover, we derive a much weaker sufficient condition than $\delta_s<0.307$ provided by \cite{Lin et al}. In the case of $D=I$, our main results return to these results in \cite{Zhang and Li 2017}. Meanwhile, the bound is sharp in the case, for more details, see \cite{Cai and Zhang 2014}.

\noindent {\bf Acknowledgments}

This work was supported by Natural Science Foundation of China (No. 61673015, 61273020) and Fundamental Research Funds for the Central
Universities(No. XDJK2015A007).

\end{document}